\documentstyle[11pt,paspconf,epsf]{article}

\begin{document}
\title{The Galaxy Luminosity and Selection Functions\\ 
of the NOG sample} 
\author{Christian Marinoni, Giuliano Giuricin, Barbara Costantini}
\affil{Dept. of Astronomy, Trieste Univ., Trieste, Italy}
\author{Pierluigi Monaco}
\affil{Institute of Astronomy, Cambridge, UK}

\begin{abstract}
 In order to map the galaxy density field on small scales in the local
 universe, we  use the  Nearby Optical  Galaxy  (NOG) sample, which is
 currently one of the largest, nearly complete, magnitude--limited ($B\leq$14
 mag), all--sky sample of nearby optical galaxies ($\sim$6400 galaxies
 with  $cz<$5500 km/s).  We  have  corrected  the  redshift--dependent
 distances of these galaxies for non-cosmological  motions by means of
 peculiar velocity  field models.

 Relying on  group assignments and  on total B magnitudes fully
 corrected  for internal   and Galactic extinctions,   we determine the
 total and morphological-type  specific luminosity functions for field
 and grouped galaxies using their locations in real distance space.
 
 The related determination of the selection function is meant to be an
 important step in recovering the galaxy density field on small scales
 from the NOG sample.  Local galaxy density parameters will be used in
 statistical studies of environmental effects on galaxy properties.       
\end{abstract}

\section{Introduction}
We  use the Nearby Optical    Galaxy (NOG) sample  to reconstruct  the
galaxy density field in the local universe. This sample is an all-sky,
magnitude-limited  sample of nearby  galaxies  (with  $cz<5500$ km/s),
which is nearly  complete down to   the limiting total  corrected blue
magnitude B=14 mag and comprises  6392 galaxies, of which 2789 objects
are members of   galaxy systems (with at  least   three members).  The
completeness level of the NOG sample limited to $|b|>15^{\circ}$ (5832
galaxies) is estimated to be $\sim$80\%.

The  redshift-dependent distances  of the  field  galaxies  and galaxy
systems  have been corrected for  non-cosmological motions by means of
peculiar  velocity field models (Marinoni  et al. 1998a). Specifically,
we  employed two independent models:  i)  a semi-linear approach which
uses a multi-attractor      model  (with Virgo,     Great   Attractor,
Perseus-Pisces  Supercluster  and Shapley  Concentration)  fitting the
Mark  III  peculiar  velocity catalog  (Willick  et  al.  1997); ii) a
modified  version  of  the  optical  cluster  3D-dipole reconstruction
scheme by Branchini \& Plionis (1996).

\section{The Total Galaxy Luminosity Function}
Adopting Turner's  (1979)    method  we  evaluate  the  total   galaxy
luminosity function  (LF) for for  field  and grouped galaxies,  using
their location in real distance space.  Since the NOG sample comprises
both bright     and   nearby galaxies,   systematic    errors   in the
determination of the LF are likely to minimized.

We find that the galaxy  LF is well  described by a Schechter function
with $\alpha\sim$-1.1, a low normalization factor $\Phi^{*}\sim$ 0.006
Mpc$^{-3}$,  and   a  particularly  bright   characteristic  magnitude
$M_B^{*}\sim$-20.7  ($H_0= 75  km^{-1}   Mpc^{-1}$)  (see Marinoni  et
al. 1998b for details).   Our $M_B^{*}$-value is brighter, on average,
by   0.4  mag than  previous   results,  because, referring  to  total
magnitudes corrected for Galactic extinction, internal extinction, and
K-dimming, better represent the galaxy light.

The exact values of the Schechter parameters of the LF slightly depend
on the adopted velocity field models (see Fig. 1), but peculiar motion
effects are  of the order of statistical  errors; at most,  they cause
variations  of 0.08 in    $\alpha$ ($\sim1\sigma$ error)  and 0.2  mag
($\sim2\sigma$ error) in $M_B^{*}$.

The presence  of galaxy  systems  in the NOG   sample does not  affect
significantly the field galaxy  LF. Environmental effects on the total
LF  are proved to be  marginal. The LF  of the  galaxy  members of the
richest systems tends to show a  slightly brighter value of $M_B^{*}$,
which gives some evidence of luminosity segregation with density.

\section{The Morphological--Type Dependence of the Luminosity Function}

We   also    evaluate the    morphological   type-specific   LFs.  The
morphological types are available for almost all NOG galaxies.

The LF  of E+S0 galaxies does  not  differ significantly from  that of
spirals.  But the  E galaxies clearly  decrease in number towards  low
luminosities (with $\alpha\sim$-0.5), whereas  the number of late-type
spirals   and  irregulars rise steeply towards   the   faint end (with
$\alpha\sim$-2.3 -- -2.4).  This behaviour hints at  an upturn  of the
total LF in the  unexplored faint end (at  $M_B>$-15).  In Fig.   2 we
show a comparison between  our type--specific LFs with those  obtained
from the  CfA2 (Marzke et al.   1994) and the Stromlo--APM (Loveday et
al. 1992; see also Driver, Windhorst \& Griffiths 1995) samples.

As regards the morphological  type--dependence of the LF, our  results
better agree with those derived from the CfA2 and SSRS2 (Marzke et al.
1998) samples than with the ones obtained from the Stromlo-APM survey.
Moreover, the  dependence   of   the  LF on  the   morphological  type
appreciably differs from    its  dependence on the    galaxy  spectral
classification as given by the LCRS (Bromley et  al. 1998) and Autofib
redshift survey (Heyl et al. 1997).

\newpage
\clearpage

\begin{figure}
\plotfiddle{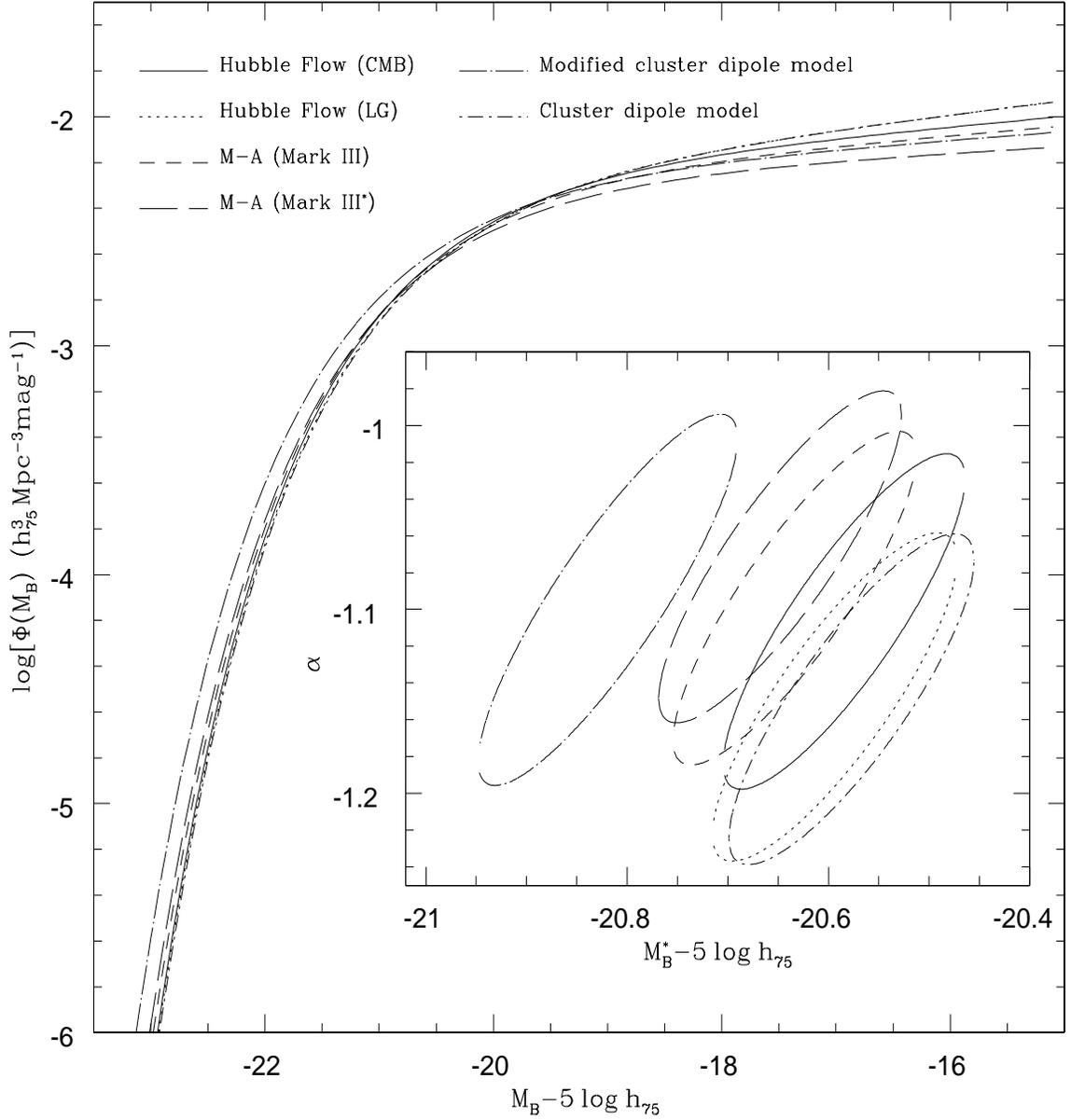}{15.5cm}{0}{80}{85}{-285}{-120}
\caption{We compare the best-fitting Schechter functions obtained from
six galaxy distance models: the Hubble  relations evaluated in the CMB
and LG   frames, the multi--attractor models fitted   on the total and
spiral (Mark III$^*$) sets of Mark III data, the cluster dipole model,
and a   modified version  of  this model    which  includes  a   local
Virgocentric  infall. The inset    shows the corresponding 1  $\sigma$
error contours for  the joint distribution of  errors of the Schechter
parameters $\alpha$ and $M_{B}^{*}$.}
\end{figure}

\newpage
\clearpage

\begin{figure}

\plotfiddle{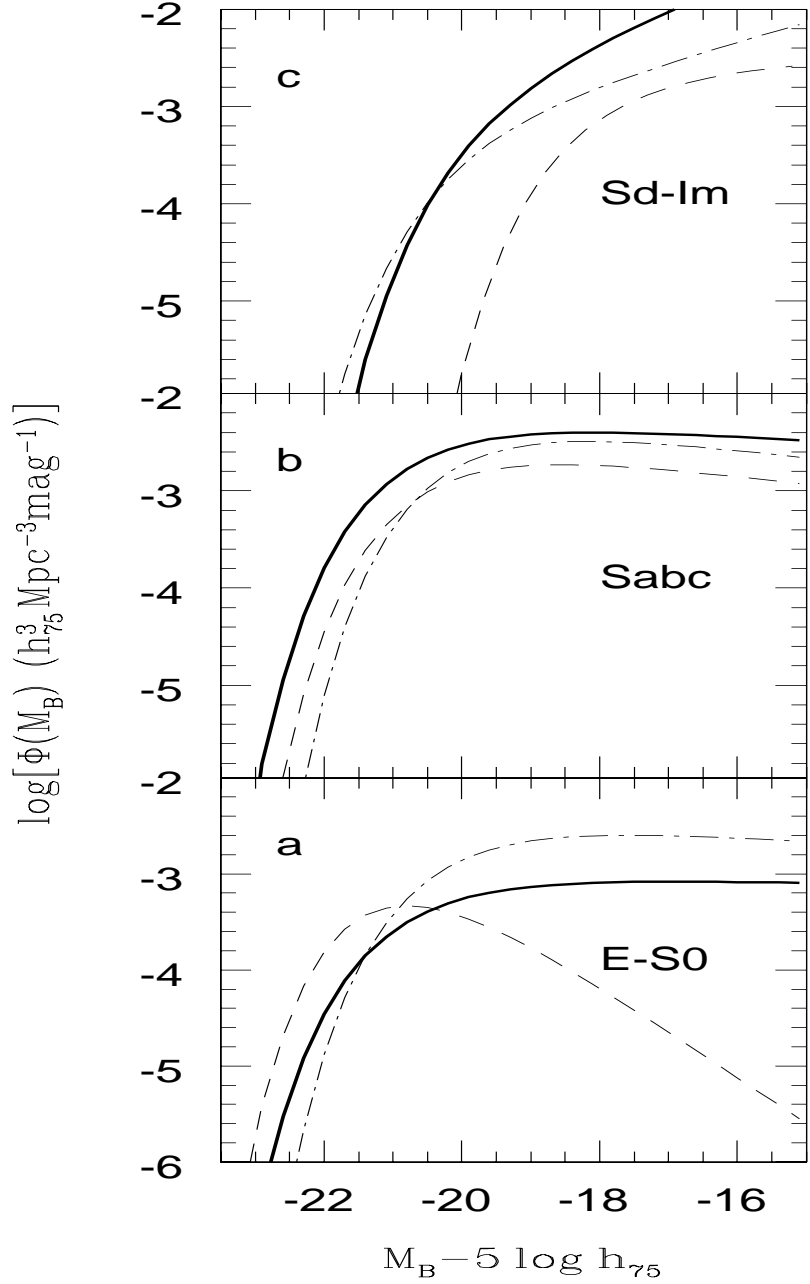}{16.5cm}{0}{140}{92}{-450}{-130}
\caption{For the E--S0 , Sabc ,  Sd--Im galaxies, we  show the NOG LFs
({\em solid line}) (with true  distances based on the multi--attractor
model fitted on Mark III data) and the corresponding LFs obtained from
the  CfA2 (Marzke  et al.  1994)  ({\em  dotted--dashed line} and  the
Stromlo--APM  (Loveday  et al.  1992; see   also Driver,  Windhorst \&
Griffiths 1995) ({\em dashed line}) amples.}
\end{figure}

\newpage
\clearpage

\section{Local Galaxy Density and Environmental Effects}

Underpredicting the observed galaxy  number counts (e.g., Ellis  1997)
at bright magnitudes ($B\sim$18 mag), where little galaxy evolution is
observed for the  bulk of  the galaxy  population, our relatively  low
local   normalization, which can not    be biased  low by  photometric
problems or by incompleteness of the  sample, suggests that the nearby
universe is underdense in galaxies (by a factor $\sim$1.5).

Although the  galaxy LF, as well as  the  intimately related selection
function of the NOG sample, appear to  be little sensitive to peculiar
motion effects,  these effects have a  quite large impact on the local
galaxy density,  especially on the  smallest  scales.

We are calculating the local galaxy density of each galaxy in terms of
the number  density   of  neighbouring galaxies.    This   is done  by
smoothing every  galaxy with a Gaussian  filter (Giuricin et al. 1993;
Monaco et al.  1994) and by  correcting the incompletion of the sample
at large distances  through the selection function  of the sample. The
main goal of this line of research is to use small-scale density
parameters   to analyze  environmental effects on    the properties of
nearby galaxies.

\end{document}